\documentclass[11pt]{article}

\usepackage{cuted} 
\usepackage{xcolor}
\usepackage{colortbl}
\usepackage[table]{xcolor}
\usepackage{amsmath}
\usepackage{url}
\usepackage{amssymb}

\usepackage{tcolorbox}
\tcbuselibrary{listings, breakable} 
\usepackage{multicol}
\usepackage{listings}
\usepackage{float}

\usepackage[table]{xcolor}  
\usepackage{booktabs} 
\usepackage{multirow}

\usepackage[final]{acl}

\usepackage{times}
\usepackage{latexsym}

\usepackage[T1]{fontenc}

\usepackage[utf8]{inputenc}

\usepackage{microtype}

\usepackage{inconsolata}

\usepackage{graphicx}

\title{TDD-Agent: Test-Driven Reasoning for Code Generation}

\author{
 \textbf{Hongyue Yu\textsuperscript{1}},
 \textbf{Kefan Li\textsuperscript{2}},
 \textbf{Jiakun Li\textsuperscript{2}},
 \textbf{Hongzheng Chai\textsuperscript{2}},
\\
 \textbf{Yuan Yuan\textsuperscript{2,3}\thanks{Corresponding authors.}},
 \textbf{Rui He\textsuperscript{4}},
 \textbf{Junyi Wei\textsuperscript{4}},
\\
\\
 \textsuperscript{1}National College for Excellent Engineers, Beihang University,
\\
 \textsuperscript{2}School of Computer Science and Engineering, Beihang University,
\\
 \textsuperscript{3}Qingdao Research Institute and Hangzhou Innovation Institute, Beihang University,
\\
 \textsuperscript{4}School of Software, Beihang University,
\\
 \texttt{Natt1e@buaa.edu.cn, kefanli@buaa.edu.cn, yuan21@buaa.edu.cn}
}

\begin{document}
\maketitle

\begin{abstract}
Large Language Models (LLMs) have achieved remarkable progress in code generation, yet ensuring correctness in complex, repository-level tasks remains challenging. Existing approaches often use generated tests as static post-hoc validators, which limits their ability to guide implementation and may introduce misleading feedback when the tests themselves are incomplete or incorrect. In this paper, we introduce TDD-Agent, which operationalizes the test-driven development paradigm for code generation. TDD-Agent first prompts the model to generate executable tests, encouraging it to clarify expected behaviors before implementation, and then performs iterative dual-track refinement over both the generated code and tests using execution feedback. 
We first isolate the effect of test-first reasoning through a prompt variant TDD-prompt on LiveCodeBench, where it consistently improves upon reasoning-based prompting baselines. Building on this finding, we evaluate the full TDD-Agent framework on RepoEval, a repository-level benchmark, and show that it consistently outperforms retrieval-based and agent-based baselines.
Additional analyses show that iterative refinement improves not only code correctness but also the effectiveness of the generated tests, yielding higher pass rates, coverage, and mutation scores,
suggesting that tests can serve as evolving reasoning artifacts rather than fixed validators. Our source code is available at \url{https://anonymous.4open.science/r/TDD-Agent-Framework-6370/}.
\end{abstract}

\section{Introduction}
\begin{figure}[t!]
    \centering
    \includegraphics[width=\linewidth]{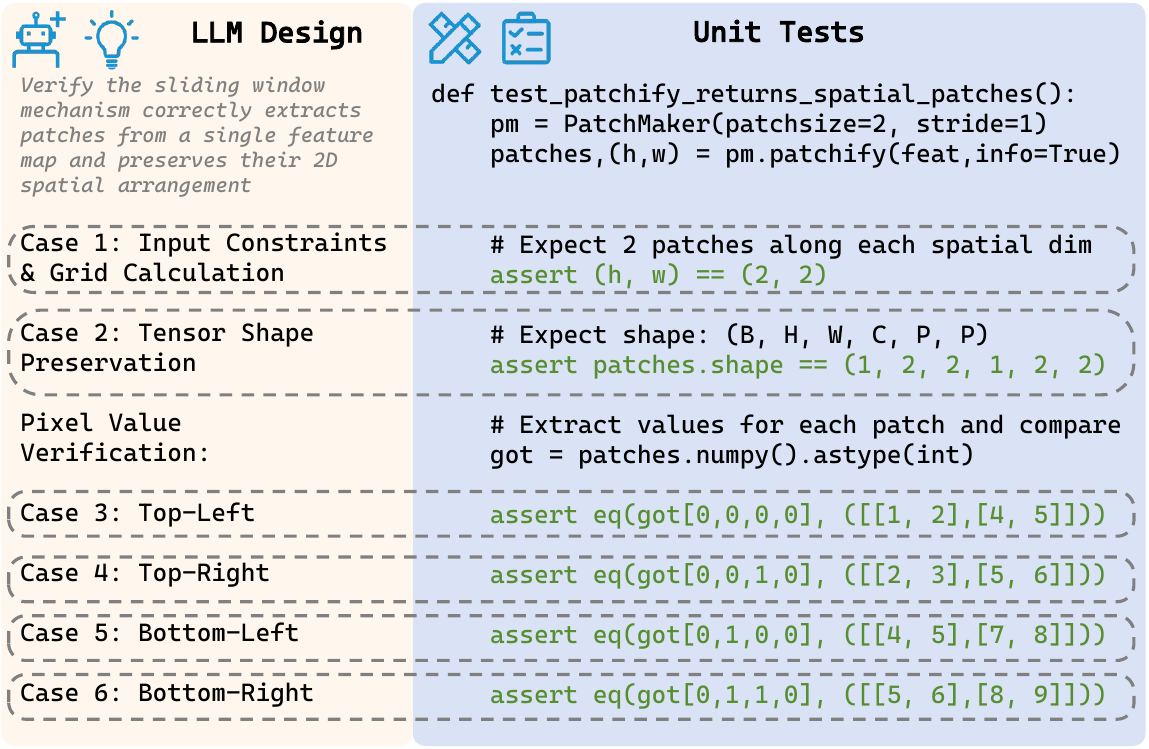}
\caption{An example demonstrating that formulating tests constitutes a form of reasoning.}
    \label{fig:example}
    \vspace{-1em}
\end{figure}

Large Language Models (LLMs) have demonstrated exceptional performance across a spectrum of software engineering tasks, particularly in code generation. The potential of these models has garnered significant attention from both academia and industry, evidenced by the rapid development of commercial foundation models such as the GPT \citep{openai2023gpt4}, Gemini \citep{team2023gemini}, and Claude \citep{anthropic2024claude} series, as well as open-source counterparts like Llama \citep{dubey2024llama}, Gemma \citep{team2024gemma}, and Qwen \citep{yang2024qwen2technicalreport}. Consequently, enhancing the efficacy of LLMs in complex coding scenarios has emerged as a critical research trajectory. 

To increase model performance, researchers have integrated software engineering methodologies into LLM-based generation. Pair Coder \citep{paircoder} employs a multi-agent framework separating high-level planning from specific implementation to improve code quality. Beyond planning, integration of software testing has proven particularly effective. 
Previous work \citep{mathews2024tdd} indicates that providing external test cases significantly boosts model performance on MBPP \citep{mbpp} and HumanEval \citep{humaneval} compared to using problem descriptions alone. 
Furthermore, AgentCoder \citep{huang2024agentcoder} employs multi-agent systems to decouple the roles of programmer and tester, demonstrating that explicit test generation can enhance code correctness.

However, the reliance on self-generated tests involves inherent risks. Recent work \citep{chen2025revisit} analyzes two paradigms: post-execution and in-execution. 
Findings reveal that post-execution debugging often suffers from test bias introduced by self-generated tests, leading to misleading feedback. 
Conversely, in-execution debugging allows LLMs to leverage intermediate states to mitigate this bias. 
Moreover, the majority of existing test-centric research is confined to function-level tasks. 
In these scenarios, generating independent functions from natural language descriptions allows for relatively trivial test synthesis (e.g., simple assert statements). 
It remains an open question whether self-testing strategies remain effective in repository-level contexts, where test generation is substantially more challenging due to dependencies and the need for environment mocking.

A limitation of prior work is that generated tests are often treated as fixed validators after they are produced.
Typically, LLMs are directed to focus solely on rectifying implementation code, without the mandate to verify or correct the validity of the tests themselves. 
We argue that this underuses the potential of testing as a structured reasoning aid. By requiring a model to formulate executable test cases before implementation, the model is encouraged to make its assumptions about inputs, outputs, edge cases, and behavioral constraints explicit. Figure \ref{fig:example} provides an example that illustrates the rationale.

Based on this insight, we instantiate TDD-Agent as a single-agent framework.  
In this paradigm, a single agent retains conversation history and is granted the autonomy to modify both the code and the tests dynamically. This approach simulates a coherent ``test-driven development (TDD)'' thought process \citep{mathews2024tdd}, allowing the model to iteratively align code and tests without the complexity overhead of coordinating multiple agents.

We empirically evaluate our approach at two levels. First, on LiveCodeBench, we use a lightweight TDD-prompt variant to isolate the effect of test-first reasoning in function-level code generation. Second, on RepoEval \citep{zhang-etal-2023-repocoder}, we evaluate the TDD-Agent framework in repository-level tasks, where repository navigation and dependency understanding are required. 

In summary, our contributions are as follows: 

\begin{itemize}
    \setlength{\itemsep}{0pt}
    \setlength{\parsep}{0pt}
    \setlength{\topsep}{0pt}
    \item We propose TDD-Agent, which adapts the test-driven development paradigm for code generation.
    \item We show that test-first generation provides an executable intermediate representation of task intent.
    \item We conduct extensive experiments on both function-level and repository-level benchmarks.
\end{itemize}

\begin{figure*}[t!]
    \centering
    \includegraphics[width=\textwidth]{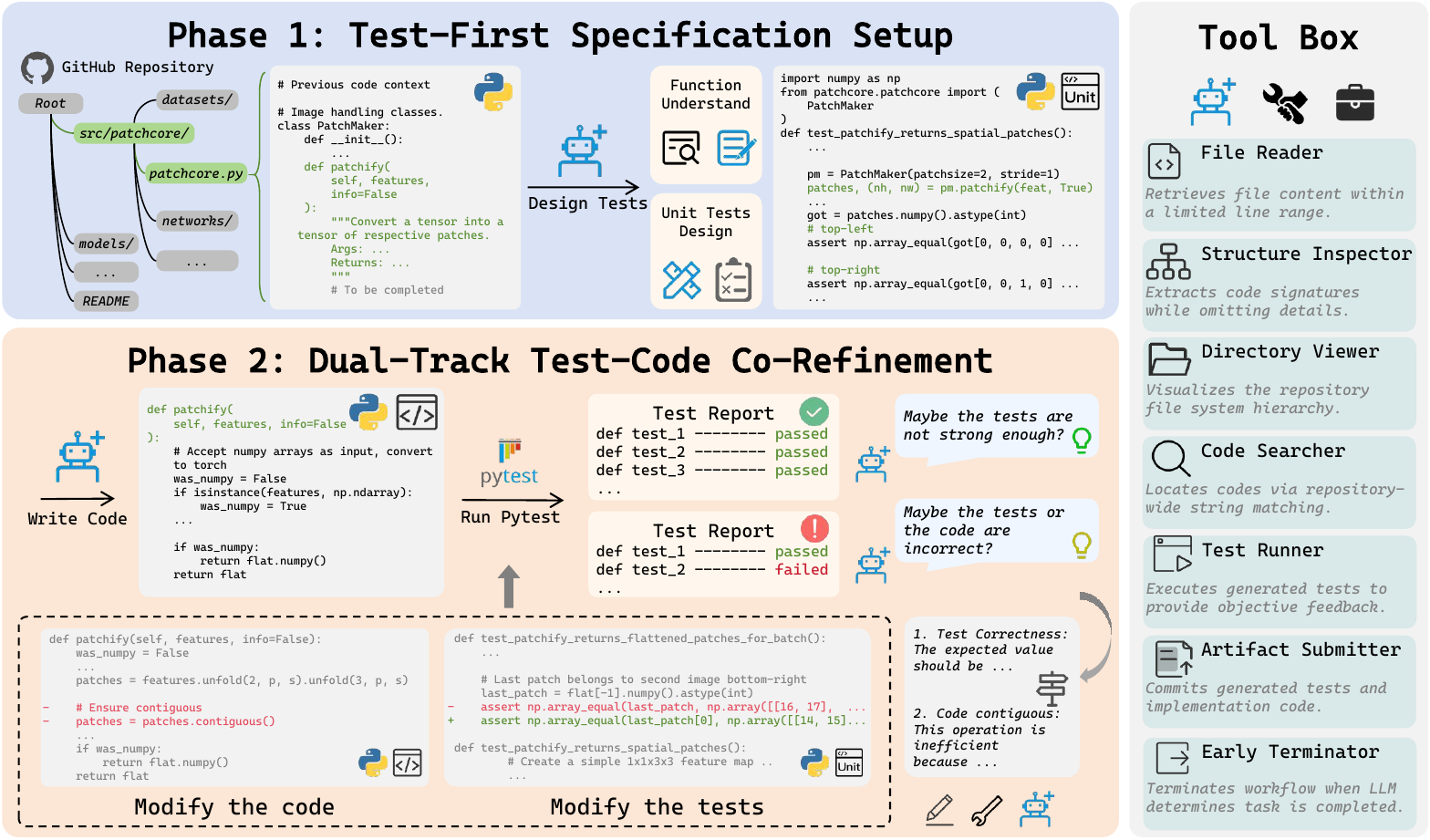}
\caption{\textbf{Overview of the TDD-Agent.} (1) The LLM first designs and generates unit tests for the target function. (2) The LLM generates the complete implementation of the target function, and through execution and refinement, iteratively refines the tests and code.}
    \label{fig:overview}
    \vspace{-1em}
\end{figure*}

\section{Methodology}

\subsection{Overall Framework}
The TDD-Agent framework operates through a two-phase workflow. 
Let $\mathcal{R}$ denote the repository context, $\mathcal{T}_{tools}$ the set of available tools, $f_{target}$ the target function to be implemented and $\mathcal{A}$ the agent. Figure \ref{fig:overview} provides a comprehensive overview of the framework.

\paragraph{Phase 1: Test-First Specification Setup}

In the initial phase, $\mathcal{A}$ acts as a test designer. It leverages tools $\mathcal{T}_{tools}$ to extract relevant context from $\mathcal{R}$ and comprehend the requirements of $f_{target}$. 
By generating an initial suite of unit tests, denoted as $U_0$, $\mathcal{A}$ is compelled to disambiguate requirements and define precise behavioral boundaries before any implementation logic is written:
\begin{equation}
    U_0 = \mathcal{A}_{design}(\mathcal{R}, f_{target} \mid \mathcal{T}_{tools}).
\end{equation}
This phase concludes with the submission of $U_0$, ensuring that the agent has established a clear, executable understanding of the task.

\paragraph{Phase 2: Dual-Track Test-Code Co-Refinement}
The second phase constitutes an iterative loop of execution and refinement. Building upon the specifications in $U_0$, $\mathcal{A}$ generates the initial implementation code $C_0$ for $f_{target}$. The implementation is then executed against $U_0$ to produce a test report $E_0$:
\begin{equation}
    E_t = \text{Execute}(C_t, U_t).
\end{equation}

Based on the execution feedback $E_t$, $\mathcal{A}$ performs a reflection process. 
\begin{itemize}
    \setlength{\itemsep}{0pt}
    \setlength{\parsep}{0pt}
    \setlength{\topsep}{0pt}
    \item If execution succeeds: $\mathcal{A}$ will receive a prompt suggesting that it consider strengthening or expanding the test suite. If $\mathcal{A}$ is confident in the correctness and robustness of $C_t$, it can choose to invoke the \texttt{Finish} tool to terminate the task early.
    \item If execution fails: $\mathcal{A}$ will receive a prompt encouraging it to analyze the cause. Unlike previous approaches that treat unit tests as immutable constraints, our framework enables the dual refinement of both \textbf{code} and \textbf{tests}.
\end{itemize} 
The state update for the next iteration is formulated as:
\begin{equation}
   (C_{t+1}, U_{t+1}) = \mathcal{A}_{reflect}(C_t, U_t, E_t, \mathcal{R})
\end{equation}
The iterative process continues until a maximum number of rounds is reached (set to 10) or when $\mathcal{A}$ invokes the \texttt{Early Terminator} tool to terminate early.

\subsection{Tool Box}
TDD-Agent is equipped with a lightweight set of tools:

\noindent\textbf{Context Inspection.}
Directory Viewer lists the repository structure, Structure Inspector extracts the skeleton of a file (e.g., class and function) with concrete implementations folded, File Reader retrieves specific code segments based on exact line numbers, Code Searcher performs lexical search using \texttt{grep} and \texttt{find}.

\noindent\textbf{Artifact Submission and Test Runner.}
The agent utilizes Artifact Submission to submit both the generated tests and implementation code. The generated tests are saved to a temporary file within the same directory as the target file, while the implementation code directly modifies the target file in place.
Test Runner strictly executes only the latest version of the generated temporary test file by running pytest.
Throughout the prediction phase, the original repository tests are neither executed nor visible to the agent.

\noindent\textbf{Early Terminator.} At the end of each iteration, specifically, upon receiving the execution results from the Test Runner, the agent can invoke this tool to exit early if it determines that the task has been completed.

\section{Experiments}
We conduct experiments on function-level and repository-level tasks using three high-performing LLMs. \textbf{GPT-5-mini} \cite{openai_gpt5_mini_2025} is a closed-source model with superior reasoning capabilities.
\textbf{DeepSeek-V3.2} \citep{deepseekv32} is a general-purpose conversational model with 671B total parameters.
\textbf{Qwen3-Coder-30B-A3B-Instruct} \citep{yang2025qwen3technicalreport} is an open-source Mixture-of-Experts model specialized for coding tasks. We access GPT and DeepSeek via APIs, while Qwen is deployed locally on one NVIDIA H20 GPU using vLLM \citep{vllm}. 
The detailed generation parameters for each model are in Appendix~\ref{appendix:parameter}.

\subsection{Function-Level Study}
We design a prompting variant TDD-prompt, which asks the LLM to formulate tests before producing the final implementation.
We conduct comparative experiments on \textbf{LiveCodeBench} \citep{jain2024livecodebenchholisticcontaminationfree} to isolate the effect of test-first reasoning. 
The detailed experiment settings are in Appendix~\ref{appendix:lcb}.

\smallskip
\noindent\textbf{Baselines.}
\textbf{CoT} \cite{cot} asks LLMs to generate natural language reasoning before generating the code.
\textbf{SCoT} \cite{SCoT} asks LLMs to use programming structures to generate structured reasoning steps before generating the code.
\textbf{Self-Planning} \cite{Self-Planning} asks LLMs to first generate a sub-task plan and then uses it to guide code generation.
\textbf{ICoT} \cite{ICoT} asks LLMs to first capture the task intention and then uses it to guide code generation.

\noindent\textbf{Results.}
To mitigate computational cost and account for inherent randomness, we generate \textbf{10} samples for each problem and measure the unbiased pass@1 metric \cite{humaneval}.
As shown in Table \ref{tab:livecodebench}, TDD-prompt outperforms all baselines on three LLMs. 
This consistent improvement demonstrates that TDD-prompt functions as a reasoning framework.

\begin{table}[t]
  \centering
  \small
  \begin{tabular}{lccc}
    \toprule
    \textbf{Method} & \textbf{GPT} & \textbf{DeepSeek} & \textbf{Qwen} \\
    \midrule
    one-shot      & 53.08 & 45.40 & 39.60 \\
    CoT-prompt    & 67.41 & \underline{67.46} & \underline{42.99} \\
    SCoT-prompt   & 66.83 & 64.20 & 40.80 \\
    Self-Planning & 68.17 & 66.70 & 42.68 \\
    ICoT-prompt   & \underline{68.48} & 62.23 & 41.88 \\
    \textbf{TDD-prompt}    & \textbf{70.04} & \textbf{67.86} & \textbf{44.87} \\
    \bottomrule
  \end{tabular}
  \caption{Pass@1 (\%) comparison on LiveCodeBench across three models. Bold and underline indicate the best and second-best performance.}
  \label{tab:livecodebench}
  \vspace{-1em}
\end{table}

\subsection{Repository-Level Experimental Setup}

\noindent\textbf{Datasets.}
We evaluate our method on the RepoEval dataset \citep{zhang-etal-2023-repocoder}, which comprises repository-level line, API, and function completion tasks. 
In this study, we focus specifically on the function completion subset, derived from eight distinct Python repositories and containing a total of 455 problems.

\smallskip

\noindent\textbf{Metrics.}
Following the protocol of RepoCoder, we employ an execution-based evaluation metric. We utilize the repository's existing unit tests to assess functional correctness. 
Specifically, we integrate the generated code into the original repository and execute the accompanying unit tests. A solution is deemed successful (\textit{Pass}) only if it passes all associated test cases; otherwise, it is marked as failed.

\noindent\textbf{Execution Environment.}
We construct a dedicated Docker environment for each repository, where all required dependencies are pre-installed and configured. Before evaluation, we verify that the original test suite provided by each repository can be executed successfully in the corresponding environment. During prediction and evaluation, each Docker container is allocated 2 CPU cores and 4 GB of memory. Each command is limited to a maximum execution time of 600 seconds. If an evaluation run exceeds the timeout, it is treated as a failure.

\subsection{Repository-Level Baselines}
\noindent\textbf{In-File Completion.}
We provide the LLM exclusively with the code context available within the current file (prefix code) and require it to complete the missing function body. 
\smallskip

\noindent\textbf{RAG.}
Following RepoCoder, we employ a sparse bag-of-words model as the retriever \citep{lu2022reaccretrievalaugmentedcodecompletion}. This model tokenizes both the query and candidate snippets to calculate similarity via the Jaccard index \citep{jaccard1912distribution}.
\smallskip

\noindent\textbf{RepoCoder.}
RepoCoder is an iterative retrieval and generation framework. We adopt exactly the same configuration specified in the original paper.
\smallskip

\noindent\textbf{mini-SWE-agent}
\cite{yang2024sweagent} is a widely adopted software engineering agent equipped with a bash tool. It achieves excellent performance on \textbf{SWE-Bench} and is recommended by its original authors. We set the maximum number of LLM calls to 100 for each problem.

\subsection{Repository-Level Main Results}

\begin{table}[t]
  \centering
  \small
  \begin{tabular}{lccc}
    \toprule
    \textbf{Method} & \textbf{GPT} & \textbf{DeepSeek} & \textbf{Qwen} \\
    \midrule
    In-File            & 44.84 & 39.34 & 32.75 \\
    RAG                & 50.55 & 46.15 & 35.16 \\
    RepoCoder          & 55.16 & 50.55 & 41.10 \\
    mini-SWE-agent     & 61.31 & 84.18 & 52.97 \\
    \textbf{TDD-Agent-iter5}  
                       & \underline{77.36} & \underline{88.13} & \underline{58.46} \\
    \textbf{TDD-Agent-iter10} 
                       & \textbf{78.24} & \textbf{90.77} & \textbf{59.34} \\
    \bottomrule
  \end{tabular}
  \caption{Performance comparison on RepoEval across three models. Numbers are presented in percentage (\%). Bold and underline indicate the best and second-best among all compared methods.}
  \label{tab:main}
  \vspace{-1em}
\end{table}

Experimental results presented in Table \ref{tab:main} demonstrate that \textbf{TDD-Agent} outperforms all baselines. The detailed results for each repository are reported in Appendix~\ref{appendix:repoeval}.

While retrieval-based methods like RAG and RepoCoder generally enhance performance by providing relevant context, their improvements are relatively modest. In contrast, agent-based methods like mini-SWE-agent and TDD-Agent achieve stronger performance gains. Notably, TDD-Agent already surpasses the mini-SWE-agent baseline by its fifth iteration, and increasing the number of iterations yields further performance gains.

For agent-based methods, efficiency is a crucial metric alongside overall performance.
We report the average token usage and LLM calls for mini-SWE-agent and TDD-Agent in Table \ref{table:main_usage}. 
We observe that at its fifth iteration, TDD-Agent exhibits comparable token consumption to mini-SWE-agent, while consistently achieving superior performance. The detailed token usage and time cost are presented in Appendix~\ref{appendix:repoeval}.
\begin{table}
  \centering
  \small
  \begin{tabular}{llccc}
    \toprule
    \textbf{Model} & \textbf{Method} & \textbf{Prompt} & \textbf{Completion} & \textbf{Calls} \\
    \midrule

    \multirow{3}{*}{GPT}
      & mini & 115.76 & 1.32 & 11.95 \\
      & TDD-iter5 & 120.48 & 1.62 & 18.62 \\
      & TDD-iter10  & 155.61 & 1.71 & 22.92\\
    \midrule

    \multirow{3}{*}{DS}
      & mini & 390.10 & 6.38 & 31.61 \\
      & TDD-iter5 & 361.64 & 9.71 & 29.39 \\
      & TDD-iter10  & 403.01 & 10.66 & 31.49 \\
    \midrule

    \multirow{3}{*}{Qwen}
      & mini & 106.83 & 2.14 & 14.81 \\
      & TDD-iter5 & 117.07 & 3.02 & 15.34 \\
      & TDD-iter10  & 129.87 & 3.22 & 17.30 \\
    \bottomrule
  \end{tabular}
  \caption{Average token usage (in thousands) and average LLM calls across different models on mini-SWE-agent and TDD-Agent.}
  \vspace{-1em}
  \label{table:main_usage}
\end{table}

\begin{figure}[t!]
    \centering
    \includegraphics[width=\linewidth]{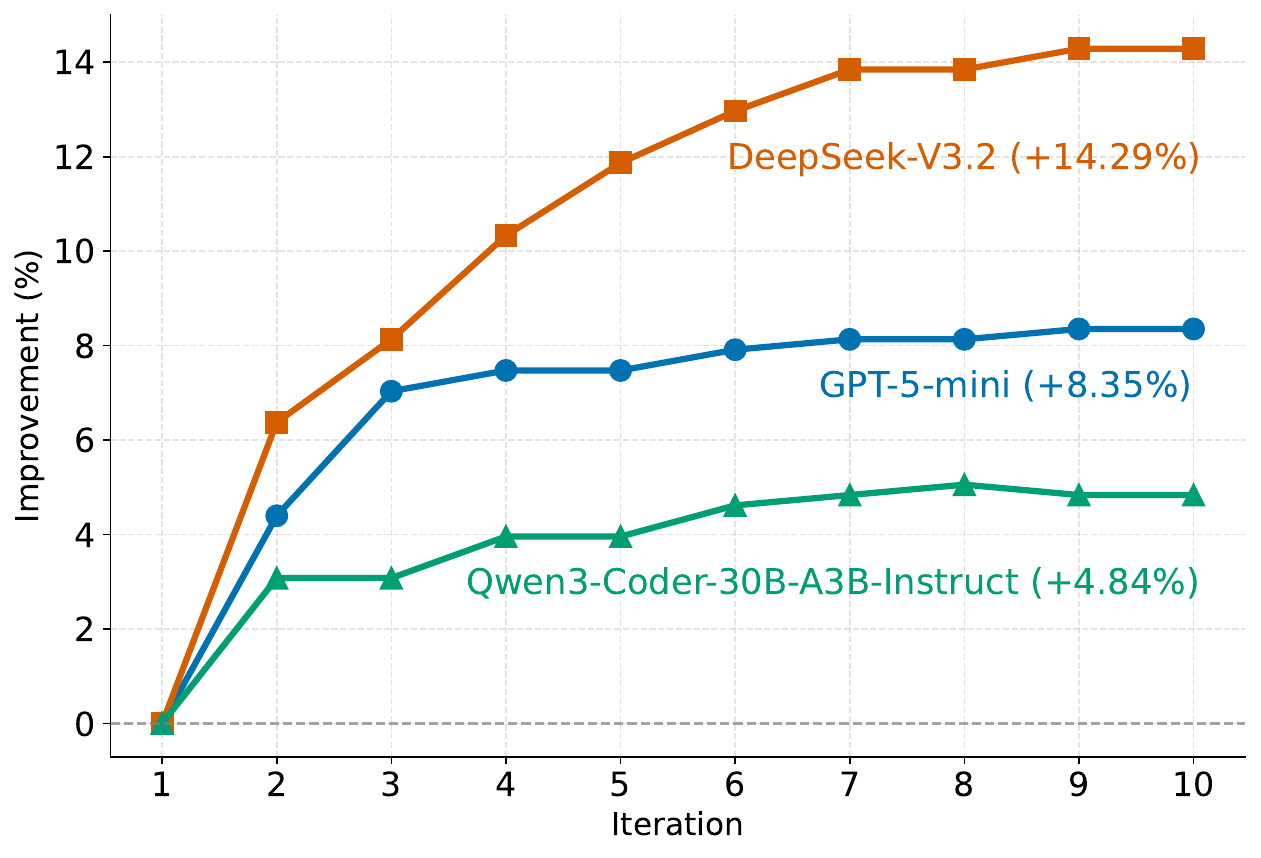}
\caption{The pass rate improvement across iterations on three models.}
    \label{fig:iteration_pr}
    \vspace{-1em}
\end{figure}

\section{Analysis}

\subsection{Early Termination}
\label{sec:confidence}

\begin{figure*}[t!]
    \centering
    \includegraphics[width=\textwidth]{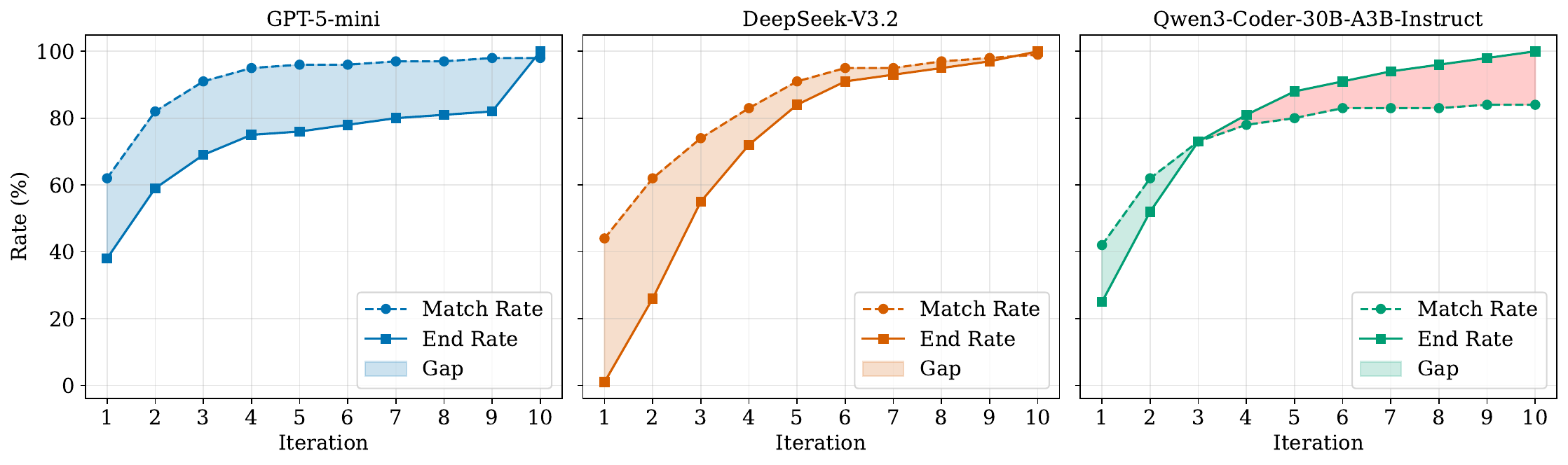}
\caption{Comparison between End Rate and Match Rate across different iteration rounds on three models. Match Rate represents the proportion of generated code that successfully passes the generated tests. End Rate indicates the proportion of tasks where the model autonomously decides to terminate.}
    \label{fig:match_vs_end}
    \vspace{-1em}
\end{figure*}

Figure~\ref{fig:match_vs_end} reports the Match Rate and End Rate across all iterations for the three models. Match Rate denotes the proportion of tasks whose current generated code passes the current generated tests, while End Rate denotes the proportion of tasks for which the model chooses to terminate at the current iteration.

The results reveal clear differences across models. In the first iteration, DeepSeek behaves the most conservatively. Although a substantial portion of its generated implementation already passes the generated tests, it rarely chooses to terminate. This indicates that DeepSeek does not simply treat self-test success as sufficient evidence of correctness. Instead, it tends to continue refining even after the current tests are passed. This conservative behavior helps explain why DeepSeek obtains the largest improvement through iterative refinement, as it keeps using additional rounds to strengthen the test suite and correct potential hidden defects.

GPT and Qwen terminate more aggressively in early iterations. For GPT, the End Rate remains consistently lower than the Match Rate, suggesting that the model generally requires self-test success before deciding to stop. This indicates a stopping heuristic: passing the generated tests is treated as an important but not always sufficient condition for termination.
Qwen exhibits a different pattern. From the middle iterations onward, its End Rate becomes higher than its Match Rate. This means for some tasks, Qwen chooses to stop even when the current implementation has not passed its generated tests.

\subsection{Improvement Across Iterations}

Figure~\ref{fig:iteration_pr} illustrates the pass rate improvement achieved by TDD-Agent across iterations for the three models. All models benefit from iterative refinement, confirming that the execution-feedback loop is effective for improving performance.

Among the three models, DeepSeek achieves the largest overall gain. This is consistent with the early-termination analysis in Section~\ref{sec:confidence}: DeepSeek is less likely to stop after the first successful self-test and instead continues to refine its artifacts. As a result, it can exploit later iterations more effectively, leading to the largest cumulative improvement.

GPT also shows a clear and stable improvement trend. Its performance increases rapidly during the first several rounds and then gradually stabilizes. This suggests that GPT can effectively use execution feedback to correct errors, while its termination behavior prevents excessive premature stopping.

Qwen obtains a smaller but still positive gain. Although the analysis in Figure~\ref{fig:match_vs_end} indicates that Qwen sometimes terminates before passing its generated tests, its performance still improves with more iterations, implying that although some tasks suffer from early stopping, the remaining active tasks can still benefit from continued refinement.

Across all three models, the improvement curves gradually converge and reach a near-plateau around iterations 6--7. This trend indicates that most useful corrections are made in the early and middle stages of the iterative process, while additional rounds provide diminishing returns. This observation suggests that a moderate iteration budget can provide a favorable trade-off between performance and computational overhead.

\subsection{Quality of Generated Tests}

A core premise of TDD-Agent is that generated tests should not be treated as static artifacts. Instead, they should evolve together with the implementation. To examine whether this property holds in practice, we analyze the quality of generated tests from three perspectives: pass rate, coverage rate and mutation score.

At each iteration, the generated test suite is executed against the ground-truth implementation. A test suite is considered passed if all its tests succeed on the ground-truth implementation. A low pass rate indicates potential issues such as incorrect assumptions or hallucinated requirements.

The coverage rate is measured using the \texttt{pytest-cov} plugin. Specifically, we execute the generated tests on the ground-truth implementation and compute line coverage for the target function:
\begin{equation}
    \text{Coverage Rate} = \frac{\text{Covered Lines}}{\text{Total Lines}}.
\end{equation}
This metric measures how thoroughly the generated tests exercise the implementation logic.

We assess mutation score using \textbf{MutPy} \cite{mutpy}. 
For each problem, we generate 20 mutants of the ground-truth implementation and execute the LLM-generated tests against them. 
The mutation score is defined as the fraction of mutants killed by the generated tests:
\begin{equation}
    \text{Mutation Score} = 
    \frac{\text{Killed Mutants}}{\text{Total Mutants}}.
\end{equation}
If the generated tests fail on the ground-truth implementation, we set the mutation score to 0.

\begin{figure*}[t!]
    \centering
    \includegraphics[width=\textwidth]{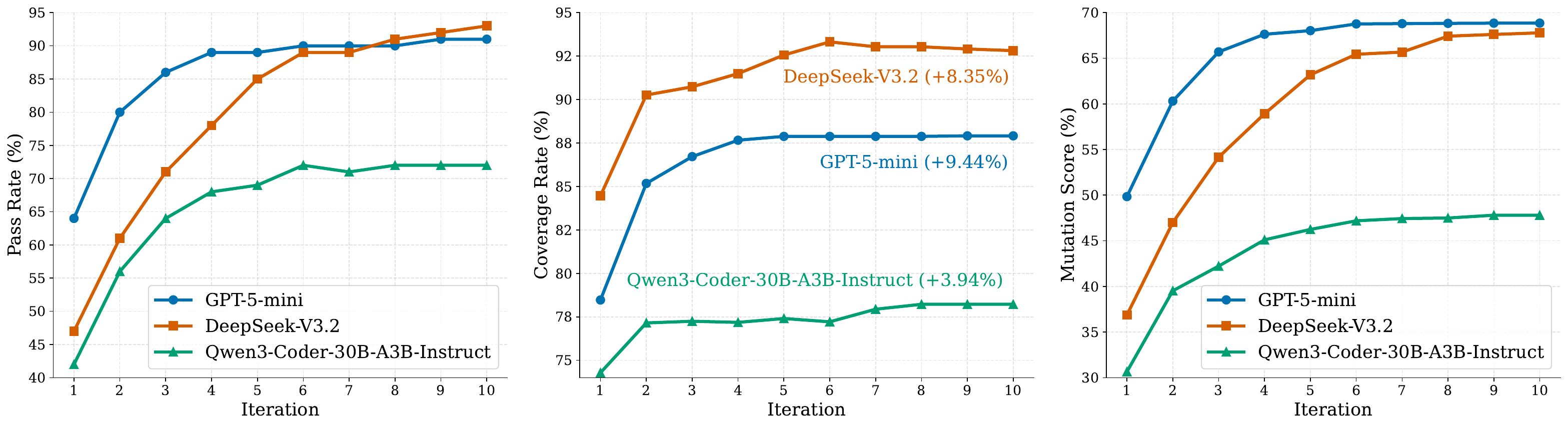}
\caption{Pass rate, coverage rate, and mutation score of generated tests over iterations for three models.}
    \label{fig:tests}
    \vspace{-1em}
\end{figure*}

Figure~\ref{fig:tests} presents the pass rate, coverage rate, and mutation score of generated tests across iterations. The results show that all three metrics generally improve as the number of iterations increases. This indicates that the iterative refinement process improves not only the generated code but also the generated tests. Through iterative refinement, the agent gradually removes invalid assertions, corrects mismatched expectations, and expands the test suite to cover more execution paths.

The tests generated by DeepSeek achieve consistently high values across all three metrics, which provides a direct explanation for its large code-generation gain. Since the generated tests become more reliable, they provide more accurate feedback, enabling the model to correct defects more effectively.
GPT also shows steady improvements, suggesting that it can consistently refine its generated tests based on execution feedback. 
Qwen starts with comparatively lower metric values and obtains a smaller gain, but still shows a positive trend. This further confirms that even when a model exhibits early-termination behavior, the dual-track refinement process can still improve the quality of its generated tests.

These findings support the central hypothesis of TDD-Agent: generated tests are most effective when they are treated not as fixed validators but as evolving reasoning artifacts.

\subsection{Ablation Study}

To disentangle the contribution of each core component in TDD-Agent, we conduct an ablation study on RepoEval.
TDD-Agent consists of two key designs: (1) test-first generation, which requires the model to construct executable specifications before implementation, and (2) dual-track refinement, which allows the model to iteratively revise both the implementation and the generated tests based on execution feedback.

We compare TDD-Agent with four variants. \textit{Vanilla} directly generates the implementation without test generation or iterative refinement. \textit{Reflect} removes test generation and performs iterative self-reflection only on the implementation.
\textit{Single-track} keeps test-first generation but freezes the generated tests during refinement, allowing only the code to be modified.
\textit{TDD-Agent-iter1} executes only the first test generation and code generation without subsequent refinement.

As shown in Table~\ref{tab:ablation}, the full TDD-Agent consistently achieves the best performance, demonstrating that both test-first reasoning and dual-track refinement are essential to the framework.
Compared with \textit{Vanilla}, \textit{TDD-Agent-iter1} achieves higher pass rates, indicating that generating tests before implementation can indeed improve code quality by encouraging the model to clarify expected behaviors in advance.
However, the improvement is limited, suggesting that a single round of test-first generation is insufficient.

The comparison between \textit{Vanilla} and \textit{Reflect} shows that iterative reflection alone can also improve performance. However, reflection alone remains limited because the model lacks concrete feedback to verify whether its revisions are actually correct.
\textit{Single-track} introduces test-first generation and therefore provides executable feedback during refinement. Nevertheless, its improvement is also limited, since the initially generated tests may contain errors or incomplete assumptions, and thus may fail to provide accurate guidance in some cases.

The gap between \textit{Single-track} and the full TDD-Agent highlights the importance of dual-track refinement.
Allowing the agent to revise both tests and code enables it to correct not only implementation errors but also flawed or insufficient tests. This refinement process helps the generated tests gradually better align with the intended behavior of the target function, thereby providing more reliable feedback for subsequent code revisions. Overall, these results confirm that test generation is most effective when it is treated not as a fixed validator, but as an evolving reasoning artifact that is refined jointly with the implementation.

\begin{table}[t]
  \centering
  \small
  \begin{tabular}{cccc}
    \toprule
    \textbf{Variant} & \textbf{GPT} & \textbf{DeepSeek} & \textbf{Qwen} \\
    \midrule
    Vanilla         & 68.35 & 73.63 & 52.97 \\
    Reflect         & \underline{72.09} & \underline{81.53} & 54.06 \\
    Single-track    & 70.11 & 79.56 & \underline{57.36} \\
    TDD-Agent-iter1 & 69.89 & 76.48 & 54.51 \\
    \textbf{TDD-Agent} & \textbf{78.24} & \textbf{90.77} & \textbf{59.34} \\
    \bottomrule
  \end{tabular}
  \caption{Ablation study on RepoEval. Results are reported as pass rate (\%). Bold and underline indicate the best and second-best performance for each model.}
  \label{tab:ablation}
  \vspace{-2em}
\end{table}

\begin{figure}[H]
    \centering
    \includegraphics[width=\linewidth]{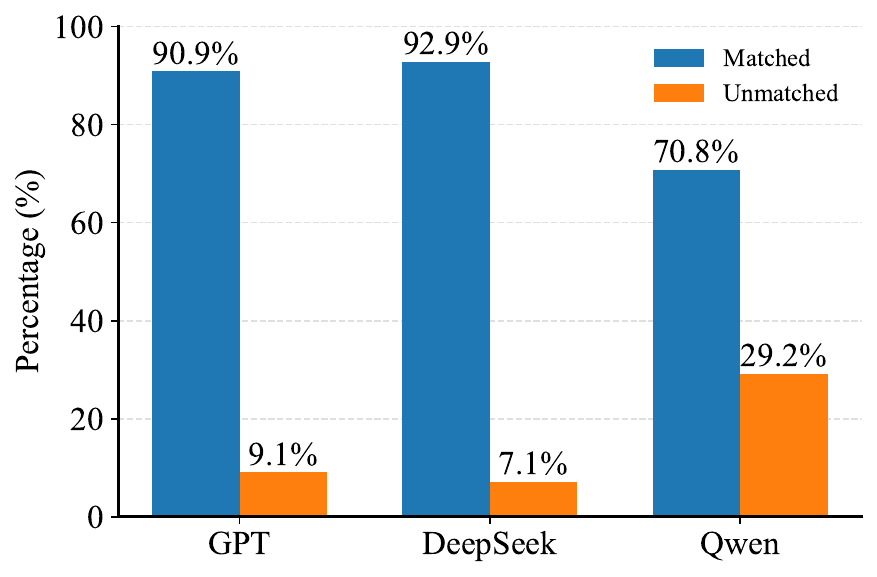}
\caption{Distribution of matched and unmatched failures across three backbone models.}
    \label{fig:match_unmatch_bar}
    \vspace{-1em}
\end{figure}

\subsection{Failure Case Analysis}

A natural concern in TDD-Agent is the \textit{credit assignment problem}. When an execution fails, the agent must determine whether the implementation is flawed or the tests are incorrect. There are cases where the agent may fix correct code to pass a hallucinated test, or conversely, discard valid tests to accommodate erroneous code. However, our statistics show that such cases account for less than 10\% of the failed cases.

We further analyze the failed cases according to whether the implementation passes the generated tests at termination.
As shown in Figure~\ref{fig:match_unmatch_bar}, most failures are matched failures, where the generated implementation passes the generated tests but fails the held-out repository tests. This indicates a form of false-positive verification: the agent has produced tests that are consistent with its implementation, yet fail to fully capture the intended behavior, acting as an incomplete or misaligned executable specification.
Since the refinement loop is driven by execution feedback from the current generated tests, defects that are not exercised by these tests may remain invisible throughout refinement. As a result, tests and code can co-evolve into an internally consistent but incomplete state: the implementation satisfies the generated tests, while both artifacts still fail to capture behaviors required by the repository oracle. This suggests that matched failures reflect not merely low test quality, but a limitation in the model’s task understanding as operationalized through test generation. More detailed statistics are reported in Appendix~\ref{appendix:repoeval}.

\section{Related Work}

\noindent \textbf{Test-Guided Code Generation.}
\textsc{TICODER} \citep{fakhoury2025ticoder} introduces an interactive workflow where models generate candidate tests to clarify user intent, while \textsc{IntUT} \citep{IntUT} uses test intentions to improve branch coverage in unit test generation. \textsc{CoCoEvo} \citep{cocoevo} co-evolves programs and test cases via genetic search, and \textsc{LLMLOOP} \citep{LLMLOOP} incorporates mutation analysis to make generated tests more challenging. 
More closely related, \textsc{TENET} \citep{TENET} leverages pre-existing repository tests for test selection, context retrieval, and feedback-guided code refinement. 
In contrast to prior work that relies on user interaction, predefined tests, or uses tests mainly as validation/refinement signals, TDD-Agent treats test generation as a reasoning step: it synthesizes executable tests before implementation and jointly refines both tests and code through execution feedback, without assuming access to predefined tests during prediction.

\noindent \textbf{Software Engineering Agents.}
\citet{yang2024sweagent} introduce \textsc{SWE-agent}, which utilizes a custom Agent-Computer Interface to prevent the model from being overwhelmed by verbose shell outputs.
\citet{zhang2024codeagent} propose \textsc{CodeAgent}, which integrates symbol navigation and testing tools, demonstrating that precise code navigation is critical for repository-level tasks. 
\citet{tao2024magis} propose \textsc{MAGIS}, a framework that assigns distinct roles to agents to simulate a real-world development lifecycle. Conversely, \citet{xia2024agentless} argue for simplicity with \textsc{Agentless}, a method that eschews agentic loops for a simpler workflow.
\textsc{OpenHands} \citep{wang2025openhands} further accelerates this field by providing a unified runtime for developing and evaluating such generalist agents.

\noindent \textbf{Retrieval-Augmented Completion.}
\textsc{RLCoder} \citep{wang2025rlcoder} employs reinforcement learning to align the retriever with code generation utility. \textsc{CoCo} \citep{zhao2025coco} parses code at multiple granularities to capture structural dependencies.
\section{Conclusion and Future Work}
In this paper, we introduced \textbf{TDD-Agent}, which operationalizes the Test-Driven Development paradigm. TDD-Agent treats test generation as a process of reasoning, compelling the model to clarify requirements and define executable boundaries prior to implementation. Through iterative refinement, our framework enables the dual refinement of both code and tests.
Experimental results demonstrate that TDD-Agent consistently outperforms baselines. Our analysis further reveals that the quality of generated tests improves concurrently with the code implementation, validating the efficacy of our dual-track refinement strategy.
For future work, we plan to extend TDD-Agent to more programming languages and further improve the reliability of self-generated tests. In particular, our failure case analysis suggests that many remaining failures arise when generated tests fail to expose hidden defects in the implementation. Future work may therefore explore stronger test-oracle construction, mutation-guided test expansion, and uncertainty-aware termination criteria to reduce false-positive verification.

\section*{Limitations}
\noindent\textbf{Scope of Evaluation Languages.} 
Our experimental validation is currently confined to Python. To adapt our framework to other languages, one only needs to replace the environment-specific components (e.g., JUnit for Java, Jest for JavaScript), while the high-level reasoning and iteration logic remain unchanged.

\noindent\textbf{Limited Repository Understanding.} 
TDD-Agent currently relies on a lightweight tool set. While these tools are sufficient to support iterative refinement, they provide limited semantic access to repository-level information.
This limitation is reflected in our failure analysis: most failures are matched failures. Such cases suggest that the dominant bottleneck is incomplete understanding of tasks and repositories. Future work could integrate stronger semantic code search, dependency-aware navigation, call-graph analysis and more informative repository-level context construction to improve the quality of generated tests and reduce false-positive verification.

\noindent\textbf{Reliability of Execution Feedback.} 
Our framework assumes that the test execution provides reliable feedback. In practice, flaky tests or nondeterministic behavior may introduce noisy signals.
Such unreliable feedback can mislead the agent into modifying correct implementations or discarding valid tests, thereby degrading the effectiveness of iterative refinement.
Future work could improve robustness by incorporating repeated test execution and flaky-test detection.

\noindent\textbf{Computational Overhead.} 
The iterative nature of TDD-Agent inherently incurs higher token consumption and latency.

\bibliography{custom}

\appendix
\section{Generation Parameter Settings}
\label{appendix:parameter}

In our experiments, we specifically utilized the \textbf{gpt-5-mini-2025-08-07} for GPT-5-mini. For Qwen3-Coder-30B-A3B-Instruct, we adopt the optimal generation parameters recommended by the official guidelines. Additionally, the temperature parameter for DeepSeek-V3.2 is set to its default value of 1.0.

\begin{table}[H]
  \centering
  \begin{tabular}{cll}
    \hline
    \textbf{Model} & \textbf{Parameter} & \textbf{Value} \\
    \hline
    \multirow{4}{*}{GPT-5-mini} 
    & temperature & 1.0 \\
    & max\_tokens & 4096 \\
    & reasoning\_effort & minimal \\
    & verbosity & low \\
    \hline
    \multirow{2}{*}{DeepSeek-V3.2} 
    & temperature & 1.0 \\
    & max\_tokens & 4096 \\
    \hline
    \multirow{5}{*}{Qwen} 
    & temperature & 0.7 \\
    & top\_p & 0.8 \\
    & top\_k & 20 \\
    & repetition\_penalty & 1.05 \\
    & max\_tokens & 4096 \\
    \hline
  \end{tabular}
  \caption{Generation parameter settings used in experiments for all models.}
  \label{tab:generation_params}
\end{table}

\section{Detailed settings of LiveCodeBench}
\label{appendix:lcb}
\smallskip
\noindent\textbf{Dataset and Evaluation Metric}

We select LeetCode-sourced problems from the most recent one-year window of \textbf{LiveCodeBench} releases (May 1, 2024 to May 1, 2025), resulting in a total of 224 problems.
We measure the effectiveness using the pass@k metric \cite{humaneval}, which provides a robust and execution-based evaluation of functional correctness. By calculating the expected probability that at least one out of $k$ generated samples passes all tests, this metric effectively mitigates the high variance introduced by the stochastic nature of LLMs during decoding. The unbiased estimator of pass@k is defined as:
\begin{equation}
pass@k = \mathop{\mathbb{E}_{problems}} \left[ 1 - \frac{\binom{n-c}{k}}{\binom{n}{k}} \right].
\end{equation}
\smallskip
\noindent\textbf{Sampling Settings}

\noindent Considering cost efficiency, we generate \textbf{10} candidates per problem. 
Following recent work \cite{ICoT}, for single-stage approaches such as one-shot and CoT, we apply the previously described sampling settings. For multi-stage approaches, including SCoT, ICoT, and Self-Planning, we use these same settings to generate 10 reasoning chains during the reasoning phase, followed by deterministic code generation at a temperature of 0.
\smallskip

\noindent\textbf{Ablation Study}

To further evaluate the effectiveness of test-first paradigm, we remove the test generation instruction from the TDD-prompt. 
The results shown in Table \ref{tab:lcb_ablation} further highlight that the test-generation-first paradigm is a critical factor, as removing the test generation instruction leads to a drop in performance.

\begin{table}[H]
    \centering
    \begin{tabular}{lccc}
    \toprule
    \textbf{Method} & \textbf{GPT} & \textbf{DeepSeek} & \textbf{Qwen} \\
    \midrule
    \textbf{TDD-prompt} & \textbf{70.04} & \textbf{67.86} & \textbf{44.87} \\
    w/o Test Generation & 68.39 & 67.28 & 43.84 \\
    \bottomrule
    \end{tabular}
    \caption{Ablation study results on LiveCodeBench across three models.}
    \label{tab:lcb_ablation}
    \vspace{-2em}
\end{table}

\begin{table*}

  \centering
    \begin{tabular}{lccccccccc}
    \toprule
    \multirow{2}{*}{\textbf{Method}}
      & \multicolumn{8}{c}{\textbf{Repo}}
      & \multirow{2}{*}{\textbf{All}} \\
    \cmidrule(lr){2-9}
      & \textbf{1} & \textbf{2} & \textbf{3} & \textbf{4}
      & \textbf{5} & \textbf{6} & \textbf{7} & \textbf{8}
      & \\
    \midrule
    \rowcolor{gray!20}
    \multicolumn{10}{c}{\textbf{GPT-5-mini}} \\
    In-File
      & 63.89 & 67.39 & 61.19 & 39.73 & 28.12 & 50.00 & 40.91 & 19.05 & 44.84\\
    RAG
      & 69.44 & 76.09 & 59.70 & 45.21 & 32.81 & 50.00 & 45.45 & 40.48 & 50.55\\
    RepoCoder
      & 72.22 & 78.26 & 70.15 & 50.68 & 39.06 & 50.00 & 40.91 & 42.86 & 55.16\\
    mini-SWE-agent
      & 80.56 & 69.57 & 68.66 & 58.22 & 46.88 & 65.63 & 63.64 & \textbf{52.38} & 61.31 \\
    TDD-Agent
      & \textbf{86.11} & \textbf{93.48} & \textbf{92.54} & \textbf{74.66} & \textbf{73.44} & \textbf{81.25} & \textbf{72.73} & \textbf{52.38} & \textbf{78.24}\\
    \midrule
    \rowcolor{gray!20}
    \multicolumn{10}{c}{\textbf{DeepSeek-V3.2}} \\
    In-File
      & 50.00 & 54.35 & 55.22 & 36.30 & 25.00 & 43.75 & 40.91 & 16.67 & 39.34 \\
    RAG
      & 52.78 & 71.74 & 64.18 & 38.36 & 28.13 & 59.38 & 40.91 & 30.95 & 46.15 \\
    RepoCoder
      & 63.89 & 63.04 & 67.16 & 47.95 & 32.81 & 59.38 & 36.36 & 35.71 & 50.55\\
    mini-SWE-agent
      & 91.67 & 93.48 & 89.55 & 84.25 & 78.13 & \textbf{93.75} & \textbf{86.36} & 59.52 & 84.18 \\
    TDD-Agent
      & \textbf{94.44} & \textbf{95.65} & \textbf{95.52} & \textbf{91.78} & \textbf{90.62} & \textbf{93.75} & 68.18 & \textbf{80.95} & \textbf{90.77}\\
    \midrule
    \rowcolor{gray!20}
    \multicolumn{10}{c}{\textbf{Qwen3-Coder-30B-A3B-Instruct}} \\
    In-File
      & 38.89 & 43.48 & 47.76 & 33.56 & 15.62 & 40.62 & 27.27 & 11.90 & 32.75 \\
    RAG
      & 52.78 & 54.35 & 40.30 & 28.08 & 17.19 & 53.12 & 18.18 & 38.10 & 35.16 \\
    RepoCoder
      & 58.33 & 56.52 & 49.25 & 34.25 & 31.25 & 53.12 & 27.27 & 33.33 & 41.10\\
    mini-SWE-agent
      & 66.67 & 58.70 & 65.67 & 54.11 & \textbf{46.88} & 59.38 & 40.91 & 21.43 & 52.97\\
    TDD-Agent
      & \textbf{77.78} & \textbf{71.74} & \textbf{73.13} & \textbf{58.21} & 40.63 & \textbf{81.25} & \textbf{45.46} & \textbf{30.95} & \textbf{59.34}\\
    \bottomrule
  \end{tabular}
  \caption{Performance comparison on RepoEval. Numbers are presented in percentage (\%), with the best performance highlighted in bold.}
  \vspace{-1em}
  \label{table:appendix_repoeval}

\end{table*}

\begin{table}[H]
  \centering
  \begin{tabular}{lcc}
    \hline
    \textbf{ID} & \textbf{Link} & \textbf{Count}\\ 
    \hline
     1   & \href{https://github.com/leopard-ai/betty}{leopard-ai/betty} & 36\\
     2   & \href{https://github.com/CarperAI/trlx}{CarperAI/trlx} & 46\\
     3   & \href{https://github.com/lucidrains/imagen-pytorch}{lucidrains/imagen-pytorch} & 67\\
     4   & \href{https://github.com/deepmind/tracr}{deepmind/tracr} & 146\\
     5   & \href{https://github.com/google/lightweight\_mmm}{google/lightweight\_mmm } & 64\\
     6   & \href{https://github.com/amazon-science/patchcore-inspection}{amazon-science/inspection} & 32\\
     7   & \href{https://github.com/facebookresearch/omnivore}{facebookresearch/omnivore} & 22\\
     8   & \href{https://github.com/maxhumber/redframes}{maxhumber/redframes} & 42\\\hline
  \end{tabular}

  \caption{Detailed information of the GitHub repositories used for RepoEval.}
  \label{tab:repos}
  \vspace{-1em}
\end{table}

\begin{table*}
  \centering
  \small
  \begin{tabular}{c|cc|cc|cccc}
    \toprule
    \multirow{2}{*}{Iter} & \multicolumn{2}{c|}{GPT-5-mini} & \multicolumn{2}{c|}{DeepSeek-V3.2} & \multicolumn{4}{c}{Qwen3-Coder-30B-A3B-Instruct} \\
    & Prompt & Completion & Prompt & Completion & Prompt & Completion & Cached & Total \\
    \midrule
    1 & 56674 & 1111 & 158337 & 3637 & 59054 & 1685 & 53510 & 7230  \\
    2 & 80363 & 1371 & 236649 & 5961 & 85475 & 2342 & 79185 & 8633  \\
    3 & 98505 & 1516 & 295616 & 7895 & 102012 & 2714 & 95369 & 9357  \\
    4 & 110656 & 1585 & 335806 & 9042 & 110371 & 2898 & 103563 & 9705  \\
    5 & 120478 & 1621 & 361637 & 9711 & 117065 & 3024 & 110155 & 9934  \\
    6 & 129245 & 1652 & 378503 & 10117 & 121636 & 3103 & 114669 & 10071  \\
    7 & 137075 & 1673 & 387326 & 10327 & 124893 & 3156 & 117883 & 10166  \\
    8 & 143976 & 1690 & 394722 & 10490 & 127595 & 3193 & 120549 & 10239 \\
    9 & 149775 & 1702 & 399190 & 10592 & 129062 & 3215 & 121998 & 10279  \\
    10 & 155611 & 1711 & 403252 & 10660 & 129866 & 3222 & 122793 & 10295 \\
    \bottomrule
  \end{tabular}
  \caption{Average token usage of TDD-Agent across iterations for three models on RepoEval.}
  \vspace{-1em}
  \label{tab:token_three_models}

\end{table*}

\begin{figure*}
    \centering
    \includegraphics[width=0.9\textwidth]{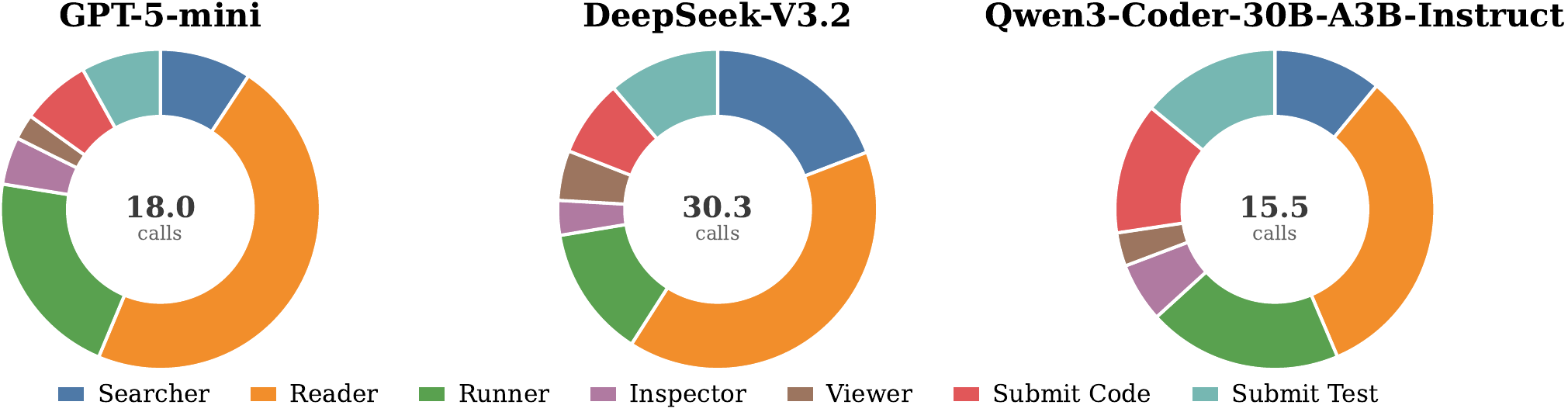}
\caption{The tool usage statistics of TDD-Agent on three models.}
    \label{fig:tool}
    \vspace{-1em}
\end{figure*}

\section{Detailed information on RepoEval}
\label{appendix:repoeval}

\noindent The detailed experimental results on RepoEval are presented in Table \ref{table:appendix_repoeval}. 
Table \ref{tab:repos} presents the information for each of the eight repositories used in RepoEval.

\noindent \textbf{Token Usage and Wall-Time}

\noindent Table \ref{tab:token_three_models} presents the detailed token usage of TDD-Agent across iterations for three models on RepoEval. Since GPT and DeepSeek are accessed via API, cached token counts cannot be precisely determined. 
Qwen is deployed locally, allowing accurate measurement of cached tokens. We only report the cached tokens for Qwen. With caching enabled, TDD-Agent’s token consumption would be further reduced.

The introduction of an iterative execution-refinement loop in TDD-Agent naturally incurs additional time overhead. Table \ref{table:main_time} reports the wall-clock time of mini-SWE-agent and TDD-Agent. We report results only on Qwen, as it is deployed locally, thereby eliminating the effects of network latency and external service load. Although additional iterations introduce extra execution time, the resulting overhead remains acceptable in light of the substantial performance improvements.

\noindent \textbf{Tool Usage Statistics}

\noindent Figure \ref{fig:tool} presents the tool usage statistics of TDD-Agent across the three models. We exclude Early Terminator from the analysis because it can be invoked at most once per task.

\begin{table}
  \centering
  \small
  \begin{tabular}{lcc}
    \toprule
    \textbf{Model} & \textbf{Method} & \textbf{Wall-clock Time(s)} \\
    \midrule
    \multirow{3}{*}{Qwen}
      & mini-SWE-agent & 26.71 \\
      & TDD-Agent-iter5   & 42.47 \\
      & TDD-Agent-iter10  &  48.43 \\
    \bottomrule
  \end{tabular}
  \caption{Average wall-clock time for mini-SWE-agent and TDD-Agent on locally deployed Qwen for one task.}
  \label{table:main_time}
  \vspace{-1.5em}
\end{table}

\noindent \textbf{Failure Case Analysis}

\noindent Qwen exhibits a relatively larger proportion of unmatched failures in Figure \ref{fig:match_unmatch_bar}. This is consistent with the analysis in Section \ref{sec:confidence}, where Qwen sometimes terminates even when its implementation does not pass the generated tests.

Table \ref{tab:credit_assignment} presents the credit assignment problem rate in failure cases across three models, with the highest value being 8.10\%, which does not exceed 10\%.

\begin{table}[H]
    \centering
    \begin{tabular}{cc}
    \toprule
    \textbf{Model} & \textbf{Credit Assignment Problem Rate} \\
    \midrule
    GPT & 3.03 \% \\
    DeepSeek & 4.76 \% \\
    Qwen    & 8.10 \% \\
    \bottomrule
    \end{tabular}
    \caption{The Credit Assignment Problem Rate in failure cases across three models.}
    \label{tab:credit_assignment}
    \vspace{-1em}
\end{table}

\begin{figure}[t]
    \centering
    \includegraphics[width=0.8\linewidth]{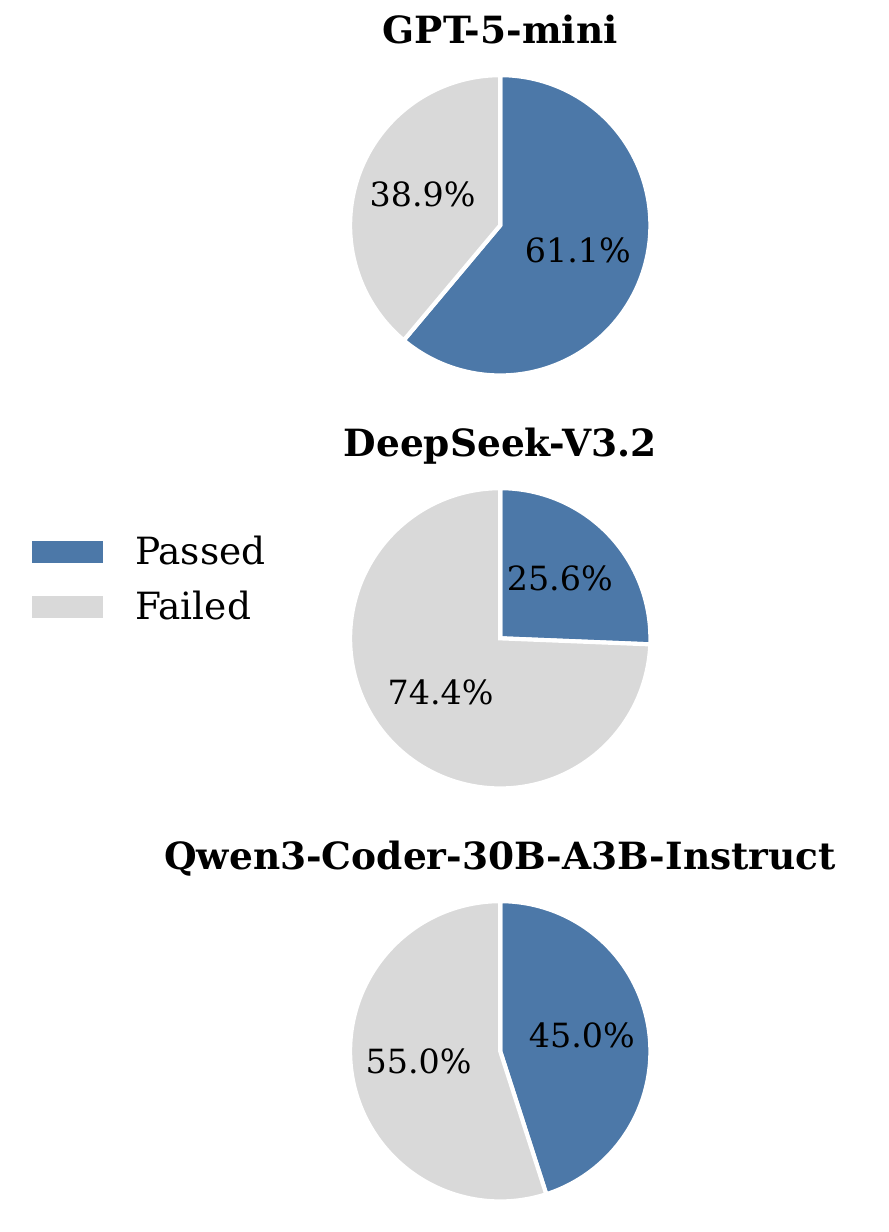}
\caption{Generated-Tests Distribution in Matched Failures.}
    \label{fig:pass_fail_pie}
    \vspace{-1em}
\end{figure}

Figure \ref{fig:pass_fail_pie} presents the distribution of generated tests on ground-truth implementations within matched failures. If generated tests pass the ground-truth implementation, it indicates that the tests are consistent with the correct solution but are not sufficiently discriminative: they fail to distinguish the incorrect implementation from the expected behavior.
In contrast, if a generated test suite fails the ground-truth implementation, this suggests that the test oracle itself is flawed, potentially due to incorrect assertions or misaligned expected outputs.
Both types of issues appear across three models, showing that matched failures arise from a mixture of weak-but-valid tests and invalid tests.
These findings suggest that future work should focus on stronger test-oracle construction, mutation-guided test expansion, and uncertainty-aware termination criteria to reduce false-positive verification.


\section{Tool Implementation Details}
\label{app:tool-implementation}

TDD-Agent is implemented as a function-calling agent whose tool invocations are parsed and compiled into a small set of command-based actions. Each tool is exposed to the model through a JSON schema specifying the tool name, description, argument types, and required fields. 
For every model response, the framework requires at least one valid tool call. 

\paragraph{Context Inspection.}
\texttt{Directory Viewer} lists the contents of a directory, optionally recursively. To avoid exposing irrelevant or excessively large metadata, it always excludes \texttt{.git}  directory, hides hidden files by default, sorts directory and file names deterministically, and caps the number of returned entries by a configurable \texttt{max\_results} parameter. 
\texttt{File Reader} reads either an entire file prefix or a user-specified inclusive line range. Line numbers are 1-based, invalid ranges are rejected, and each call is capped at 200 lines to prevent the model from receiving excessively long file contents in a single observation. 
\texttt{Structure Inspector} is handled by the runtime and returns a lightweight structural summary of a Python file, including classes, function or method signatures, and line numbers, while omitting full implementations.
\texttt{Code Searcher} provides repository-wide lexical search while avoiding unrestricted shell access. The model may issue raw Linux-style search commands, but the parser enforces three restrictions before execution. First, dangerous shell metacharacters and constructs, including command separators, redirection operators, subshells, and backticks, are rejected. Second, the pipe symbol is the only allowed command-composition operator. Third, after tokenization with \texttt{shlex}, every command appearing either at the beginning of the command or immediately after a pipe must belong to a fixed allowlist:
\texttt{find}, \texttt{grep}, \texttt{egrep}, \texttt{fgrep}, \texttt{xargs}, \texttt{head}, \texttt{tail}, \texttt{cat}, \texttt{less}, \texttt{wc}, \texttt{sort}, \texttt{uniq}, \texttt{awk}, and \texttt{cut}.
If a working directory is specified, the framework changes into that directory using shell-quoted paths before executing the validated command. This gives the agent enough flexibility for code search while preventing arbitrary command execution.

\paragraph{Artifact Submission and Test Runner.}
\texttt{Artifact Submitter} records the tests and implementation generated by the LLM. Specifically, submitted tests are written directly to \texttt{test\_by\_agent.py}, whereas submitted implementation code is used to replace the corresponding target code in the original repository file.
\texttt{Test Runner} takes no arguments and executes only the most recently submitted generated test file, using \texttt{pytest}. During prediction, the repository's original hidden evaluation tests are never exposed to the agent and are not executable through \texttt{Test Runner}. They are used only after prediction for final evaluation.
After each call to \texttt{Test Runner}, the resulting pytest report is appended to the conversation context and used as execution feedback for the next refinement step. The model may then revise the tests, revise the implementation, inspect more context, or call \texttt{Early Terminator}. The framework terminates when \texttt{Early Terminator} is called or when the predefined maximum number of refinement rounds is reached. In our main experiments, this maximum is set to 10.

\section{All Prompts Used in Experiments}
Table~\ref{table:prompt-lcb} and Table~\ref{table:prompt-lcb-ablation} present the prompts used in the function-level experiments and ablation studies.

For repository-level experiments, Table~\ref{table:prompt-repoeval} presents the prompt for the full TDD-Agent setting, while Table~\ref{table:prompt-repoeval-vanilla}, Table~\ref{table:prompt-repoeval-reflect}, and Table~\ref{table:prompt-repoeval-single-track} present the prompts for the Vanilla, Reflect, and Single-track variants, respectively.

\begin{table*}[t]
  \centering
  \caption{Prompt for TDD-prompt on LiveCodeBench.}
  \label{table:prompt-lcb}

  \small
  \begin{tabular}{>{\raggedright\arraybackslash}p{0.95\textwidth}}
    \toprule
    \textbf{\# System} \\
    You are an expert Python programmer. Given a programming problem, you must follow a strict thinking process before writing the final implementation. \\
    \midrule
    \textbf{\# User} \\
        \# Problem Description: \\
        \\
        \{description\} \\
        \\
        ```python \\
        \{starter\_code\} \\
        ``` \\
        \\
        Please solve the problem by strictly following this process:\\
        
        1. Overview: Identify the core logic, I/O format, constraints and edge cases.\\
        2. Tests: Design 3-5 concrete test cases covering basic and edge scenarios. \\
          - Use small-scale inputs to avoid miscalculation.\\
          - You MUST briefly derive the correct expected output step-by-step.\\
        3. Algorithmic Strategy: Briefly outline the optimal approach and its time/space complexity.\\
        4. Implementation: Write the final Python code inside a ```python ``` block.\\
    \bottomrule
  \end{tabular}

\end{table*}

\begin{table*}[t]
  \centering
  \caption{Prompt for Ablation Study on LiveCodeBench.}
  \label{table:prompt-lcb-ablation}
  \small
  \begin{tabular}{>{\raggedright\arraybackslash}p{0.95\textwidth}}
    \toprule
    \textbf{\# System} \\
    You are an expert Python programmer. Given a programming problem, you must follow a strict thinking process before writing the final implementation. \\
    \midrule
    \textbf{\# User} \\
        \# Problem Description: \\
        \\
        \{prompt\} \\
        \\
        ```python \\
        \{starter\_code\} \\
        ``` \\
        \\
        Please solve the problem by strictly following this process:\\
        
        1. Overview: Identify the core logic, I/O format, constraints and edge cases.\\
        2. Algorithmic Strategy: Briefly outline the optimal approach and its time/space complexity.\\
        3. Implementation: Write the final Python code inside a ```python ``` block.\\
    \bottomrule
  \end{tabular}

\end{table*}

\begin{table*}[t]
  \centering
  \caption{Prompt for TDD-Agent on RepoEval.}
  \label{table:prompt-repoeval}
  \small
  \begin{tabular}{>{\raggedright\arraybackslash}p{0.95\textwidth}}
    \toprule
    \textbf{\# System} \\
        You're a software engineer working inside an existing Python codebase rooted at /testbed.\\
        Your task is to write test cases and complete the implementation of a target function whose implementation is currently missing.
        
        You can use these tools:\\
        1. \texttt{read\_files}\\
        2. \texttt{search\_command}\\
        3. \texttt{inspect\_structure}\\
        4. \texttt{list\_files}\\
        5. \texttt{submit\_implementation}\\
        6. \texttt{submit\_tests}\\
        7. \texttt{run\_tests}\\
        8. \texttt{finish}\\
        
        Each of your response SHOULD include reasoning text explaining what you're doing\\
        Each of your response MUST include AT LEAST ONE tool call.\\
    \midrule
    \textbf{\# User} \\
        \# Task Instructions\\
        \\
        \#\# Workflow\\
        \\
        1. Inspect the target file and related code with \texttt{inspect\_structure}, \texttt{read\_files}, \texttt{search\_command} and \texttt{list\_files}\\
        
        2. Create pytest test cases for the target function and submit with \texttt{submit\_tests}\\
        \quad- Focus on the canonical usage rather than obscure edge cases or stress testing\\
        \quad- Your submitted tests will be saved as \texttt{test\_by\_agent.py} in the same directory as the target function file path\\
        \quad- It will be later executed from the project root with \texttt{python -m pytest <path\_to\_target\_dir>/test\_by\_agent.py}\\
        
        3. Complete the implementation of the target function and submit with \texttt{submit\_implementation}\\
        \quad- Submit ONLY the implementation of the target function including the function signature but WITHOUT any imports or code outside the target function definition\\
        \quad- DO NOT include anything else like any imports or contextual code\\
        \quad- \texttt{submit\_implementation} will directly OVERWRITE the target function in the original file with your submitted code\\
        
        4. After submitting both your implementation and tests, validate with \texttt{run\_tests}\\
        \quad- Before you call \texttt{run\_tests}, make sure both your implementation and tests are submitted\\
        \quad- Calling \texttt{run\_tests} will execute your latest submitted tests and implementation by running \texttt{python -m pytest <path\_to\_target\_dir>/test\_by\_agent.py} command from the project root\\
        
        5. Iteratively analyze the execution results, refine your tests and implementation, re-submit and re-run\\
        
        6. Only when you are completely confident about both your tests and your implementation should you finish your task with \texttt{finish}\\
    \midrule
    \textbf{\# Execution Succeeds Prompt} \\
    Your last execution was successful.\\
    Now Reflect on :\\
    \quad-whether the tests you wrote are sufficiently comprehensive and whether they cover all expected behaviors of the target function.\\
    \quad-whether your implementation has any remaining issues and whether it fully realizes the expected behavior of the target function.\\
    Only when you are completely confident about your implementation and tests, should you call \texttt{finish} to end your task.\\

    Note:\\
      \quad1. \texttt{run\_tests} ONLY executes your latest submitted implementation and test cases.\\
      \quad2. Running the project's pre-existing test suite is not allowed. Do not attempt to use \texttt{run\_tests} to execute the repository's native test suite.\\
      \quad3. Before you call \texttt{run\_tests} again, please ensure that you have resubmitted at least one of tests and implementation.\\
      \quad4. Rerunning \texttt{run\_tests} without any new submissions is meaningless and will only yield identical results.\\
      \quad5. Do not resubmit identical code before calling \texttt{finish}. The system will automatically save the exact version from your most recent \texttt{run\_tests} execution.\\
    \midrule
    \textbf{\# Execution Fails Prompt} \\
    Your last execution failed.\\
    Carefully analyze the reasons for failure:\\
    \quad-Is the implementation incorrect or incomplete?\\
    \quad-Are the tests you wrote insufficient or incorrect?\\
    Continue calling tools to refine your implementation and tests, submit and validate with \texttt{run\_tests}.\\
    \bottomrule
  \end{tabular}
  \vspace{-2em}
\end{table*}

\begin{table*}[t]
  \centering
  \caption{Prompt for Ablation Study on RepoEval: Vanilla.}
  \label{table:prompt-repoeval-vanilla}
  \small
  \begin{tabular}{>{\raggedright\arraybackslash}p{0.95\textwidth}}
    \toprule
    \textbf{\# System} \\
        You're a software engineer working inside an existing Python codebase rooted at /testbed.\\
        Your task is to write test cases and complete the implementation of a target function whose implementation is currently missing.
        
        You can use these tools:\\
        1. \texttt{read\_files}\\
        2. \texttt{search\_command}\\
        3. \texttt{inspect\_structure}\\
        4. \texttt{list\_files}\\
        5. \texttt{submit\_implementation}\\
        6. \texttt{finish}\\
        
        Each of your response SHOULD include reasoning text explaining what you're doing\\
        Each of your response MUST include AT LEAST ONE tool call.\\
    \midrule
    \textbf{\# User} \\
        \# Task Instructions\\
        \\
        \#\# Workflow\\
        \\
        1. Inspect the target file and related code with \texttt{inspect\_structure}, \texttt{read\_files}, \texttt{search\_command} and \texttt{list\_files}\\
        
        2. Complete the implementation of the target function and submit with \texttt{submit\_implementation}\\
        \quad- Submit ONLY the implementation of the target function including the function signature but WITHOUT any imports or code outside the target function definition\\
        \quad- DO NOT include anything else like any imports or contextual code\\
        \quad- \texttt{submit\_implementation} will directly OVERWRITE the target function in the original file with your submitted code\\
        
        3. Finish your task with \texttt{finish}\\
    \bottomrule
  \end{tabular}
  \vspace{-2em}
\end{table*}

\begin{table*}[t]
  \centering
  \caption{Prompt for Ablation Study on RepoEval: Reflect.}
  \label{table:prompt-repoeval-reflect}
  \small
  \begin{tabular}{>{\raggedright\arraybackslash}p{0.95\textwidth}}
    \toprule
    \textbf{\# System} \\
        You're a software engineer working inside an existing Python codebase rooted at /testbed.\\
        Your task is to write test cases and complete the implementation of a target function whose implementation is currently missing.
        
        You can use these tools:\\
        1. \texttt{read\_files}\\
        2. \texttt{search\_command}\\
        3. \texttt{inspect\_structure}\\
        4. \texttt{list\_files}\\
        5. \texttt{submit\_implementation}\\
        6. \texttt{finish}\\
        
        Each of your response SHOULD include reasoning text explaining what you're doing\\
        Each of your response MUST include AT LEAST ONE tool call.\\
    \midrule
    \textbf{\# User} \\
        \# Task Instructions\\
        \\
        \#\# Workflow\\
        \\
        1. Inspect the target file and related code with \texttt{inspect\_structure}, \texttt{read\_files}, \texttt{search\_command} and \texttt{list\_files}\\
        
        2. Complete the implementation of the target function and submit with \texttt{submit\_implementation}\\
        \quad- Submit ONLY the implementation of the target function including the function signature but WITHOUT any imports or code outside the target function definition\\
        \quad- DO NOT include anything else like any imports or contextual code\\
        \quad- \texttt{submit\_implementation} will directly OVERWRITE the target function in the original file with your submitted code\\
        
        3. Iteratively reflect on your implementation, analyze whether there are any remaining issues, refine your implementation and re-submit\\
        
        4. Only when you are completely confident about your implementation should you finish your task with \texttt{finish}\\
    \bottomrule
  \end{tabular}
  \vspace{-2em}
\end{table*}

\begin{table*}[t]
  \centering
  \caption{Prompt for Ablation Study on RepoEval: Single-track.}
  \label{table:prompt-repoeval-single-track}
  \small
  \begin{tabular}{>{\raggedright\arraybackslash}p{0.95\textwidth}}
    \toprule
    \textbf{\# System} \\
        You're a software engineer working inside an existing Python codebase rooted at /testbed.\\
        Your task is to write test cases and complete the implementation of a target function whose implementation is currently missing.
        
        You can use these tools:\\
        1. \texttt{read\_files}\\
        2. \texttt{search\_command}\\
        3. \texttt{inspect\_structure}\\
        4. \texttt{list\_files}\\
        5. \texttt{submit\_implementation}\\
        6. \texttt{submit\_tests}\\
        7. \texttt{run\_tests}\\
        8. \texttt{finish}\\
        
        Each of your response SHOULD include reasoning text explaining what you're doing\\
        Each of your response MUST include AT LEAST ONE tool call.\\
    \midrule
    \textbf{\# User} \\
        \# Task Instructions\\
        \\
        \#\# Workflow\\
        \\
        1. Inspect the target file and related code with \texttt{inspect\_structure}, \texttt{read\_files}, \texttt{search\_command} and \texttt{list\_files}\\
        
        2. Create pytest test cases for the target function and submit with \texttt{submit\_tests}\\
        \quad- Focus on the canonical usage rather than obscure edge cases or stress testing\\
        \quad- Your submitted tests will be saved as \texttt{test\_by\_agent.py} in the same directory as the target function file path\\
        \quad- It will be later executed from the project root with \texttt{python -m pytest <path\_to\_target\_dir>/test\_by\_agent.py}\\
        
        3. Complete the implementation of the target function and submit with \texttt{submit\_implementation}\\
        \quad- Submit ONLY the implementation of the target function including the function signature but WITHOUT any imports or code outside the target function definition\\
        \quad- DO NOT include anything else like any imports or contextual code\\
        \quad- \texttt{submit\_implementation} will directly OVERWRITE the target function in the original file with your submitted code\\
        
        4. After submitting both your implementation and tests, validate with \texttt{run\_tests}\\
        \quad- Before you call \texttt{run\_tests}, make sure both your implementation and tests are submitted\\
        \quad- Calling \texttt{run\_tests} will execute your submitted tests and your latest submitted implementation by running \texttt{python -m pytest <path\_to\_target\_dir>/test\_by\_agent.py} command from the project root\\
        
        5. Iteratively analyze the execution results, refine your implementation, re-submit and re-run\\
        
        6. Only when you are completely confident about both your implementation should you finish your task with \texttt{finish}\\
    \midrule
    \textbf{\# Execution Succeeds Prompt} \\
    Your last execution was successful.\\
    Now Reflect on whether your implementation has any remaining issues and whether it fully realizes the expected behavior of the target function.\\
    Only when you are completely confident about your implementation, should you call \texttt{finish} to end your task.\\
    Note:\\
      \quad1. \texttt{run\_tests} ONLY executes your latest submitted implementation.\\
      \quad2. Running the project's pre-existing test suite is not allowed. Do not attempt to use \texttt{run\_tests} to execute the repository's native test suite.\\
      \quad3. Before you call \texttt{run\_tests} again, please ensure that you have resubmitted your implementation.\\
      \quad4. Rerunning \texttt{run\_tests} without any new submissions is meaningless and will only yield identical results.\\
      \quad5. Do not resubmit identical code before calling \texttt{finish}. The system will automatically save the exact version from your most recent \texttt{run\_tests} execution.\\
    \midrule
    \textbf{\# Execution Fails Prompt} \\
    Your last execution failed.\\
    Carefully analyze the reasons for failure:\\
    \quad-Is the implementation incorrect or incomplete?\\
    Continue calling tools to refine your implementation, submit and validate with \texttt{run\_tests}.\\
    \bottomrule
  \end{tabular}
  \vspace{-2em}
\end{table*}

\section{Case Study}
Table~\ref{table:case-lcb-problem} and Table~\ref{table:case-lcb-response} present an example problem and the corresponding model response for TDD-prompt in the function-level experiment.

Table~\ref{table:case-repoeval-test} and Table~\ref{table:case-repoeval-code} present an example of the generated tests and implementation produced by TDD-Agent in the repository-level experiment.

\begin{table*}[t]
  \centering
  \caption{Case Study: An Example Problem from LiveCodeBench and TDD-Prompt’s Final Answer.}
  \label{table:case-lcb-problem}
  \small
  \begin{tabular}{>{\raggedright\arraybackslash}p{0.95\textwidth}}
    \toprule
\textbf{Problem Description: } \\
\\
You are given an integer array enemyEnergies denoting the energy values of various enemies.\\
You are also given an integer currentEnergy denoting the amount of energy you have initially.\\
You start with 0 points, and all the enemies are unmarked initially.\\
You can perform either of the following operations zero or multiple times to gain points:\\
\\
Choose an unmarked enemy, i, such that currentEnergy >= enemyEnergies[i]. By choosing this option:\\
\\
You gain 1 point.\\
Your energy is reduced by the enemy's energy, i.e. currentEnergy = currentEnergy - enemyEnergies[i].\\
If you have at least 1 point, you can choose an unmarked enemy, i. By choosing this option:\\
Your energy increases by the enemy's energy, i.e. currentEnergy = currentEnergy + enemyEnergies[i].\\
The enemy i is marked.\\
Return an integer denoting the maximum points you can get in the end by optimally performing operations.\\
\\
Example 1:\\
Input: enemyEnergies = [3,2,2], currentEnergy = 2\\
Output: 3\\
Explanation:\\
The following operations can be performed to get 3 points, which is the maximum:\\
First operation on enemy 1: points increases by 1, and currentEnergy decreases by 2. So, points = 1, and currentEnergy = 0.\\
Second operation on enemy 0: currentEnergy increases by 3, and enemy 0 is marked. So, points = 1, currentEnergy = 3, and marked enemies = [0].\\
First operation on enemy 2: points increases by 1, and currentEnergy decreases by 2. So, points = 2, currentEnergy = 1, and marked enemies = [0].\\
Second operation on enemy 2: currentEnergy increases by 2, and enemy 2 is marked. So, points = 2, currentEnergy = 3, and marked enemies = [0, 2].\\
First operation on enemy 1: points increases by 1, and currentEnergy decreases by 2. So, points = 3, currentEnergy = 1, and marked enemies = [0, 2].\\
\\
Example 2:\\
Input: enemyEnergies = [2], currentEnergy = 10\\
Output: 5\\
Explanation: \\
Performing the first operation 5 times on enemy 0 results in the maximum number of points.\\
 \\
Constraints:\\
1 <= enemyEnergies\.length <= $10^5$\\
1 <= enemyEnergies[i] <= $10^9$\\
0 <= currentEnergy <= $10^9$ \\ 
    \midrule
\textbf{Final Answer: } 
\begin{lstlisting}[language=Python, showstringspaces=false]
class Solution:
    def maximumPoints(self, enemyEnergies: List[int], currentEnergy: int) :
        if not enemyEnergies:
            return 0
        min_e = min(enemyEnergies)
        if currentEnergy < min_e:
            return 0
        total_sum = sum(enemyEnergies)
        return (currentEnergy + total_sum - min_e) // min_e
\end{lstlisting}  \\
    \bottomrule
  \end{tabular}
  
\end{table*}

\begin{table*}[t]
  \centering
  \caption{Case Study: A Response Example Generated by TDD-Prompt.}
  \label{table:case-lcb-response}
  \small
  \begin{tabular}{>{\raggedright\arraybackslash}p{0.95\textwidth}}
    \toprule
\begin{lstlisting}[language=Python, showstringspaces=false]
1. Overview
- Core logic: You can repeatedly "attack" an unmarked enemy i (if currentEnergy >= e_i) to gain +1 point and consume e_i energy, and you can "mark" any unmarked enemy (only if you already have >=1 point) to gain e_i energy and prevent future attacks on that enemy. Marking does not cost points. The optimal strategy is:
  - Use one attack on a cheapest enemy (minimal e) to get the required 1 point to unlock marking.
  - Then mark all other enemies to collect their energies (marking never costs points, only prevents attacking those particular enemies later).
  - Finally, spend the accumulated energy attacking the cheapest enemy as many times as possible.
- I/O: Implement Solution.maximumPoints(enemyEnergies: List[int], currentEnergy: int) -> int.
- Constraints: n up to 1e5, energies and currentEnergy up to 1e9. Use O(n) time and 64-bit integers when summing.
- Edge cases:
  - If currentEnergy < min(enemyEnergies): can't get the first point -> answer 0.
  - Single enemy case handled naturally.
  - Large sums: use Python ints (unbounded)

2. Tests (each with step-by-step expected output)
Test A: enemyEnergies = [3, 2, 2], currentEnergy = 2
- min = 2, sum = 7. current >= min so:
  - Attack min once: energy = 2 - 2 = 0, points = 1.
  - Mark the other enemies (2 and 3): energy = 0 + (2 + 3) = 5.
  - Attack min (2) floor(5 / 2) = 2 more times.
  - Total points = 1 + 2 = 3.
- Expected output: 3
Test B: enemyEnergies = [2], currentEnergy = 10
- min = 2, sum = 2. current >= min:
  - Attack once: energy = 8, points = 1. No others to mark.
  - Attack min floor(8 / 2) = 4 more times.
  - Total = 1 + 4 = 5.
- Expected output: 5
Test C: enemyEnergies = [5, 10, 3], currentEnergy = 3
- min = 3, sum = 18. current >= min:
  - Attack min once: energy = 0, points = 1.
  - Mark others (5 + 10): energy = 0 + 15 = 15.
  - Attack min (3) floor(15 / 3) = 5 more times.
  - Total = 1 + 5 = 6.
- Expected output: 6
Test D (edge: can't start): enemyEnergies = [4, 7, 9], currentEnergy = 3
- min = 4. current < min => cannot perform any attack and cannot mark.
- Expected output: 0

3. Algorithmic Strategy
- Let min_e = min(enemyEnergies) and sum_all = sum(enemyEnergies).
- If currentEnergy < min_e: return 0.
- Perform one attack on the cheapest enemy (cost min_e) to get the required 1 point.
- Mark all the other enemies and collect their energies: net energy after that = currentEnergy - min_e + (sum_all - min_e) = currentEnergy + sum_all - 2 * min_e.
- Use that energy to attack the cheapest enemy as many times as possible: floor(net_energy / min_e).
- Total points = 1 + floor(net_energy / min_e). This simplifies to (currentEnergy + sum_all - min_e) // min_e.
- Time complexity: O(n). Space complexity: O(1) extra space.

4. Implementation
```python
class Solution:
    def maximumPoints(self, enemyEnergies: List[int], currentEnergy: int) :
        ...
```
\end{lstlisting}
    \\ 
    \bottomrule
  \end{tabular}
  \vspace{-2em}
\end{table*}

\begin{table*}[t]
  \centering
  \caption{Case Study: An Example of Generated Tests by TDD-Agent.}
  \label{table:case-repoeval-test}
  \small
  \begin{tabular}{>{\raggedright\arraybackslash}p{0.95\textwidth}}
    \toprule
\begin{lstlisting}[language=Python, showstringspaces=false]
import jax
import jax.numpy as jnp
import numpy as np
import numpyro
from numpyro import handlers
import numpyro.distributions as dist
from lightweight_mmm.core import priors
from lightweight_mmm.core.transformations import saturation

def test_hill_custom_prior_half_max_effective_concentration():
    """Test that custom prior for half_max_effective_concentration is used."""
    expected_value1, expected_value2 = 5.2, 7.56
    custom_priors = {priors.HALF_MAX_EFFECTIVE_CONCENTRATION:
            dist.Kumaraswamy(
                concentration1=expected_value1, concentration0=expected_value2)}
    data = jnp.ones((10, 5))
    trace_handler = handlers.trace(handlers.seed(saturation.hill, rng_seed=0))
    trace = trace_handler.get_trace(data=data, custom_priors=custom_priors)
    prior_name = priors.HALF_MAX_EFFECTIVE_CONCENTRATION
    used_distribution = trace[prior_name]["fn"]
    used_distribution = used_distribution.base_dist
    assert isinstance(used_distribution, dist.Kumaraswamy)
    assert used_distribution.concentration0 == expected_value2
    assert used_distribution.concentration1 == expected_value1
    
def test_hill_custom_prior_slope():
    """Test that custom prior for slope is used."""
    expected_value1, expected_value2 = 3.0, 4.0
    custom_priors = {priors.SLOPE:
            dist.Kumaraswamy(
                concentration1=expected_value1, concentration0=expected_value2)}
    data = jnp.ones((10, 5))
    trace_handler = handlers.trace(handlers.seed(saturation.hill, rng_seed=0))
    trace = trace_handler.get_trace(data=data, custom_priors=custom_priors)
    prior_name = priors.SLOPE
    used_distribution = trace[prior_name]["fn"]
    used_distribution = used_distribution.base_dist
    assert isinstance(used_distribution, dist.Kumaraswamy)
    assert used_distribution.concentration0 == expected_value2
    assert used_distribution.concentration1 == expected_value1
    
def test_hill_with_zero_data():
    """Test hill function with zero input data"""
    data = jnp.zeros((10, 3))
    def mock_model():
        output = saturation.hill(data=data, custom_priors={})
        return numpyro.deterministic("output", output)
    trace_handler = handlers.trace(handlers.seed(mock_model, rng_seed=0))
    trace = trace_handler.get_trace()
    output = trace["output"]["value"]
    # With zero data, hill function should return zeros
    assert jnp.allclose(output, 0, atol=1e-7)

def test_hill_shape_3d():
    """Test that hill function handles 3D input correctly."""
    data = jnp.ones((100, 3, 5))
    def mock_model():
        output = saturation.hill(data=data, custom_priors={})
        return numpyro.deterministic("output", output)
    trace_handler = handlers.trace(handlers.seed(mock_model, rng_seed=0))
    trace = trace_handler.get_trace()
    output = trace["output"]["value"]
    assert output.shape == data.shape
\end{lstlisting}
    \\ 
    \bottomrule
  \end{tabular}
  \vspace{-2em}
\end{table*}

\begin{table*}[t]
  \centering
  \caption{Case Study: An Example of Generated Implementation by TDD-Agent.}
  \label{table:case-repoeval-code}
  \small
  \begin{tabular}{>{\raggedright\arraybackslash}p{0.95\textwidth}}
    \toprule
\begin{lstlisting}[language=Python, showstringspaces=false]
def hill(
    data: jnp.ndarray,
    custom_priors: Mapping[str, dist.Distribution],
    *,
    prefix: str = "",
) -> jnp.ndarray:
  """Transforms the input data with the adstock and hill functions.

  Args:
    data: Media data to be transformed. It is expected to have 2 dims for
      national models and 3 for geo models.
    custom_priors: The custom priors we want the model to take instead of the
      default ones. The possible names of parameters for hill_adstock and
      exponent are "lag_weight", "half_max_effective_concentration" and "slope".
    prefix: Prefix to use in the variable name for Numpyro.

  Returns:
    The transformed media data.
  """
  default_priors = priors.get_default_priors()

  with numpyro.plate(
      name=f"{prefix}{priors.HALF_MAX_EFFECTIVE_CONCENTRATION}_plate", 
      size=data.shape[1]):
    half_max_effective_concentration = numpyro.sample(
        name=f"{prefix}{priors.HALF_MAX_EFFECTIVE_CONCENTRATION}",
        fn=custom_priors.get(priors.HALF_MAX_EFFECTIVE_CONCENTRATION,
                             default_priors[priors.HALF_MAX_EFFECTIVE_CONCENTRATION]))

  with numpyro.plate(
      name=f"{prefix}{priors.SLOPE}_plate", 
      size=data.shape[1]):
    slope = numpyro.sample(
        name=f"{prefix}{priors.SLOPE}",
        fn=custom_priors.get(priors.SLOPE, default_priors[priors.SLOPE]))

  if data.ndim == 3:
    half_max_effective_concentration = jnp.expand_dims(half_max_effective_concentration, axis=-1)
    slope = jnp.expand_dims(slope, axis=-1)

  return _hill(
      data=data,
      half_max_effective_concentration=half_max_effective_concentration,
      slope=slope)
\end{lstlisting}
    \\ 
    \bottomrule
  \end{tabular}
  \vspace{-2em}
\end{table*}

\end{document}